\documentclass[preprint,groupedaddress,footinbib,amssymb]{revtex4}
\usepackage{graphicx}
\usepackage[usenames]{color}
\usepackage{amstext}
\usepackage{amsmath}
\usepackage{amssymb}
\usepackage{amsfonts}

\begin{document}

\title{Spin Revolution}

\author{E. Y. Vedmedenko$^{1}$ and R. Wiesendanger $^{1}$}
\email[corresp.\ author: ]{vedmeden@physnet.uni-hamburg.de}
\affiliation{$^{1}$University of Hamburg, Institute for Applied Physics, Jungiusstr. 11a, 20355 Hamburg}

\date{\today}

\begin{abstract}
The classical laws of physics are usually invariant under time reversal. Here, we reveal a novel class of magnetomechanical effects rigorously breaking time-reversal symmetry. The effect is based on the mechanical rotation of a hard magnet around its magnetization axis in the presence of friction and an external magnetic field, which we call spin revolution. The physical reason for time-reversal symmetry breaking is the spin revolution and not the dissipation. The time-reversal symmetry breaking leads to a variety of unexpected effects including upward propulsion on vertical surfaces defying gravity as well as magnetic gyroscopic motion that is perpendicular to the applied force. In contrast to the spin, the angular momentum of spin revolution ${\overrightarrow{L}}_{{\rm R}}$ can be parallel or antiparallel to the equilibrium magnetization ${\overrightarrow{M}}_{{\rm eq}}$ . The spin revolution emerges spontaneously, without external rotations, and offers various applications in areas such as magnetism, robotics and energy harvesting.
\end{abstract}

\maketitle

Symmetry breaking leads to fascinating effects across sciences, from the appearance of spontaneous magnetization to exciting properties of two-dimensional layered material systems \cite{Du:2021,Smej:CA2020}. Here, we reveal a novel magnetomechanical effect rigorously breaking time-reversal symmetry. State-of-the-art \textit{gyroscopic} effects involve the motion of spinning objects. A spinning axis can be defined by its mechanical angular momentum $\vec{L}_{{\rm s}} $ and velocity $\vec{\Omega }_{{\rm s}} $. A spinning object can be controlled or manipulated by another external rotation with angular velocity $\vec{\Omega }$ to align $\vec{\Omega }_{{\rm s}} $ and $\vec{\Omega }$ due to the Coriolis force as shown in Fig.1a \cite{LL}.
\begin{figure}
\includegraphics[width=0.93\linewidth]{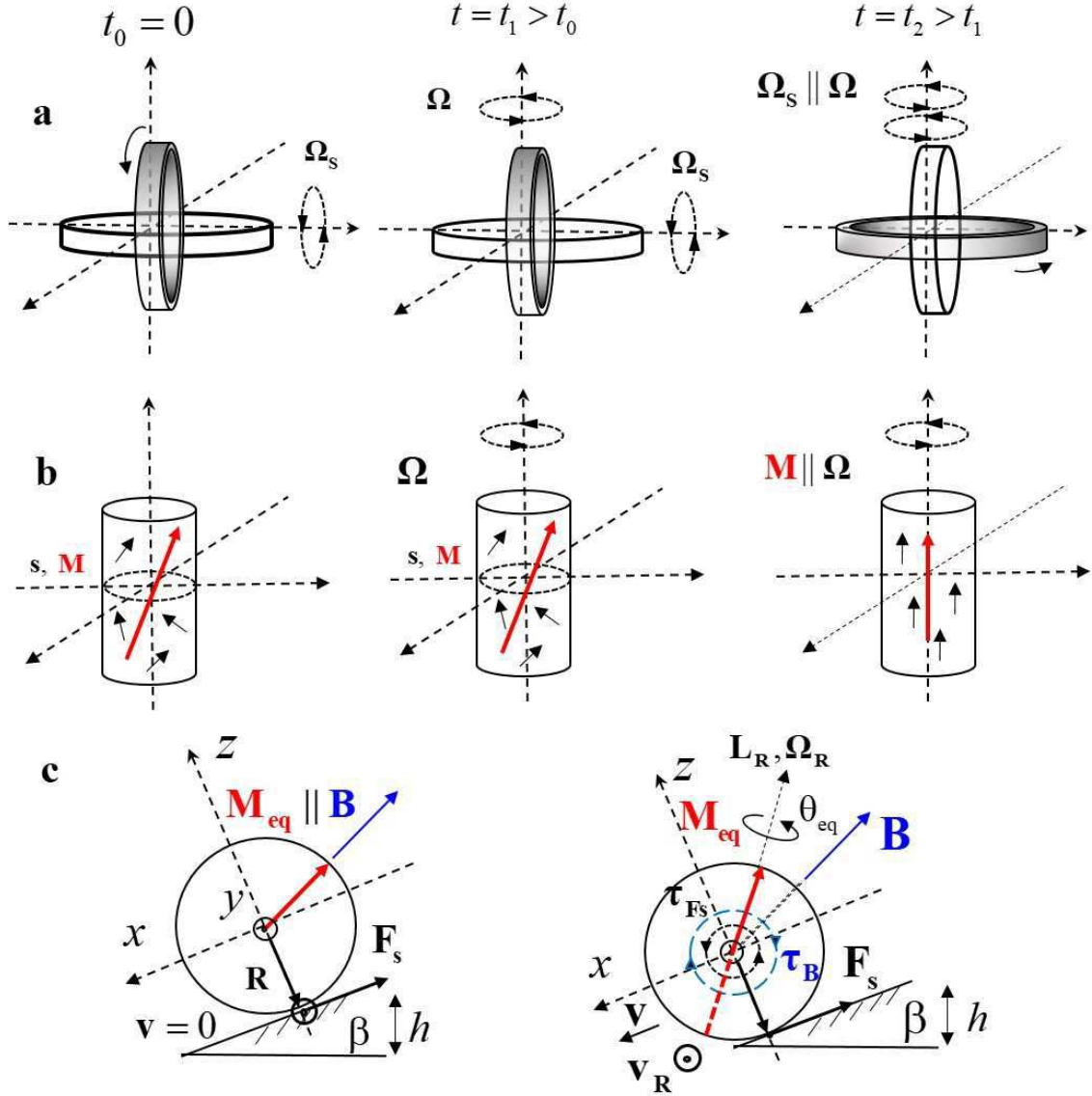}
	\caption{ \textbf{(a)} Schematic representation of a mechanical gyroscope, which can be controlled by an external rotation $\Omega $ to align the spinning axis $\Omega _{{\rm s}} $ with $\Omega $. \textbf{(b)} Schematic representation of a gyromagnetic effect, in which the magnetization can be controlled by external rotation $\Omega $ to align the magnetization $M$ with $\Omega $. \textbf{(c)} Schematic side view of a spin revolution effect. Initially, the magnetized sphere is at rest and $M||B$. When the sphere starts to roll down an incline, the magnetization departs from its initial orientation, relaxes to a direction ensuring minimal total torque ${\rm {\mathcal T}}_{{\rm c.m.}} \to \min $, and starts to revolve with $\Omega _{{\rm R}} $. Vectors are represented by bold letters for clarity.}\label{fig:F1}
\end{figure}
State-of-the-art \textit{gyromagnetic} effects are based on the motion of spinning magnetic objects. In this case, the magnetization $\vec{M}$ stemming from the spin angular momentum can be controlled or manipulated by an external rotation $\vec{\Omega }$ to align $\vec{M}$ and $\vec{\Omega }$ via the spin-rotation coupling as shown in Fig.1 (b) \cite{Matsuo:PRL11,Matsuo:JPSJ17,Heims,Barnett,EinsteinHaas}.
\begin{figure}
\includegraphics[width=0.93\linewidth]{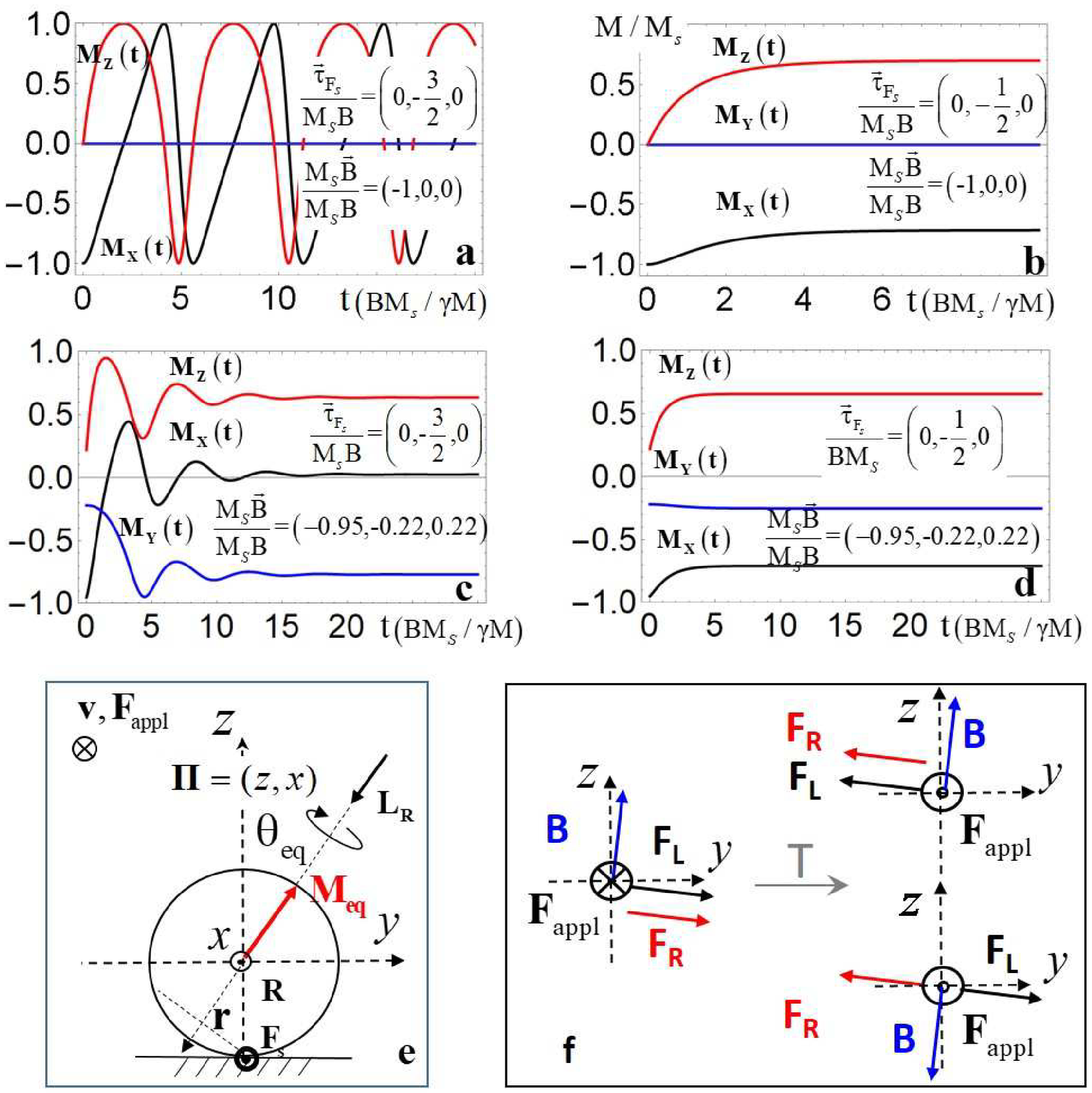}
	\caption{\textbf{Dynamics of a revolving-up magnet.}Initially, the magnetization $\vec{M}$ is parallel to a magnetic field $\vec{B}$. Next, a mechanical torque $\vec{\tau }_{{\rm F}s} $ is applied and $\vec{M}(t)$ evolves for \textbf{(a)} $(\vec{v},\vec{B})\in \Pi =\left(\hat{z},\hat{x}\right)$ and $|\vec{\tau }_{{\rm F}s} |>|\vec{\tau }_{{\rm B}}^{{\rm m}ax} |$; \textbf{(b)} $(\vec{v},\vec{B})\in \Pi $ and $|\vec{\tau }_{{\rm F}s} |\le |\vec{\tau }_{{\rm B}}^{{\rm m}ax} |$; \textbf{(c)} $(\vec{v},\vec{B})\rlap{$/$}\in \Pi $ and $|\vec{\tau }_{{\rm F}s} |>|\vec{\tau }_{{\rm B}}^{{\rm m}ax} |$; \textbf{(d)} $(\vec{v},\vec{B})\rlap{$/$}\in \Pi $ and $|\vec{\tau }_{{\rm F}s} |<|\vec{\tau }_{{\rm B}}^{{\rm m}ax} |$; \textbf{(e)} Schematical representation of geometrical axes, forces and angular momentum acting on a rolling magnetic sphere with inclined axis. All definitions correspond to the text; \textbf{(f)} Comparison of the Lorentz force for a positive charge and the force due to the SR under the local ($\vec{F}_{{\rm appl}} $ reverses, $\vec{B}$ remains unchanged) and global  ($\vec{F}_{{\rm appl}}$ reverses, $\vec{B}$ reverses) time reversal.}\label{fig:F2}
\end{figure}
In all these cases, an object subject to manipulation is initially spinning around a well-defined axis $\vec{\Omega }_{{\rm s}} $ in the laboratory frame. The magnetomechanical effect introduced here concerns a hard magnetic object (conducting or insulating) with magnetization $\vec{M}$ that does not spin initially, but rather rests at a particular position (see Fig.1c). Several torques, including a gravitational and a magnetic torque, are acting on the object and we are interested in the characteristics of the resulting movement of such an object (e.g. a sphere). In a first approximation, we neglect all effects of moving electric charges or electric fields due to the time-dependent magnetization because of their weakness. Furthermore, we are interested in the regime of rolling without slipping. Our analysis shows that, when a net torque about an object's rolling axis is minimized $\vec{{\rm T} }_{{\rm c.m.}} =\sum _{i=1}^{N}\vec{\tau }_{i}  \to \min $, the object spins up with an angular momentum $\vec{L}_{{\rm R}} $, pointing in a direction which differs from those of the magnetic field, the magnetic torque, the rolling axis, and the net torque about the rolling axis, and starts to move perpendicularly to an applied force with a velocity $\vec{v}_{{\rm R}}$. We denote this combination of spontaneous rotation and translational movement as spin revolution (SR). In contrast to known effects, the SR emerges spontaneously, without application of any external rotation about $\vec{L}_{{\rm R}} $. In contrast to the electron spin, which is antiparallel to its magnetic moment, $\vec{L}_{{\rm R}} $ can be parallel or antiparallel to an equilibrium magnetization orientation $\vec{M}_{{\rm eq}} $. The SR breaks the time-reversal symmetry of the moving magnet and leads to circular trajectories as well as to vertical propulsion defying gravity.  The key ingredient for this counterintuitive motion is the conservation of $\vec{M}_{{\rm eq}}$ leading to an additional torque about the $\vec{M}_{{\rm eq}}$ axis. The time-reversal symmetry breaking paves the way for an effective interconversion of translational and rotational motion and, by that means, to numerous applications in mechanics, robotics, energy harvesting and magnetism. For example, SR allows the development of rope-less, rail-less and hydraulics-less elevators, linear motors or angle gears. It can also be applied to the controlled rotational and translational motion of magnetic particles.
\begin{figure}
\includegraphics[width=0.83\linewidth]{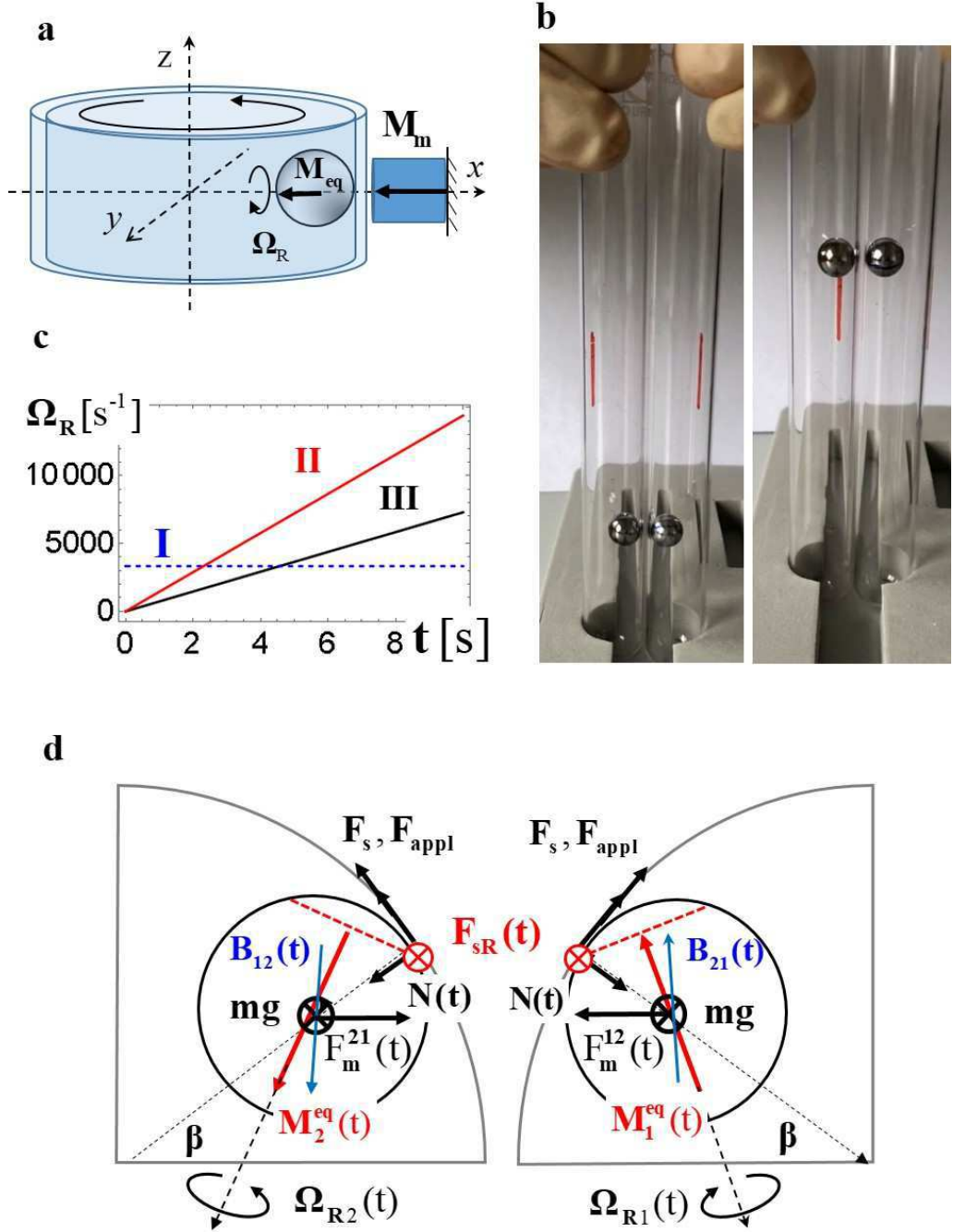}
	\caption{\textbf{Different embodiments of the spin revolution.} \textbf{(a)} Reciprocal embodiment leading to revolution of a sphere without its lateral displacement; \textbf{(b)} Numerically calculated $\vec{\Omega }_{{\rm R}} $ for a NiCoB sphere with $m=3\cdot 10^{-4} $ kg, $R=3\cdot 10^{-3} $ m and $M_{{\rm s}} =0.5{\rm A}\cdot m^{2} $, and friction coefficient $k=0.1$ for \textbf{ I} embodiment of \textbf{(a)} with $v=5$ m/s, \textbf{ II} rolling down an incline with $\beta =\pi /10$ and $\vec{B}=(-0.95,-0.22,0.22)10^{-5} $ T; and \textbf{ III} rolling down an incline with $\beta =\pi /10$ and $\vec{B}=(0,1,0)10^{-5} $ T; \textbf{(c) }Two subsequent snap-shots of SR embodiment consisting of two magnetic spheres (NiCoB, $m=5\cdot 10^{-4} $ kg, $R=3\cdot 10^{-3} $m and $M_{{\rm s}} =0.5{\rm A}\cdot m^{2} $) inside of two non-magnetic tubes. When the tubes are rotating around their vertical axes, the spheres move upwards; \textbf{(e)} Top-view of the set-up (c). Red arrows indicate the magnetic moments, red dashed lines show the rolling planes, red circles indicate the orientation of rolling friction. Blue arrows represent magnetic fields.}\label{fig:F3}
\end{figure}

To illustrate the nature of the effect and to make it more transparent, we start with a standard problem in rotational kinematics considering a sphere rolling down an inclined plane and later switch to more complicated cases. A rolling motion can be represented as a combination of a rotation about a rolling axis and its translation. A rotational torque arises from an instantaneous static friction force$\; \vec{F}_{{\rm s}} $ and equals $\vec{\tau }_{Fs} =\vec{R}\; \times \vec{F}_{{\rm s}} $  with $\vec{R}\; $ being the vector connecting the center-of-mass (c.m.) and a contact point. In the coordinate system connected with the contact point (see Fig.1c) the rolling axis coincides with $\vec{\tau }_{Fs} =(0,\frac{2}{7} mgR{\rm \; sin}\beta ,0)=(0,\tau _{{\rm Fs}}^{{\rm y}} ,0)$, with $\beta$ being the inclination angle (see \cite{Supplemental}, part A), \textit{m} the mass and \textit{g} the gravitational acceleration. It is well known that a homogeneously magnetized solid sphere is equivalent to a point dipole placed at its center \cite{ChemPhysChem,Edwards:EJP}. So, if a sphere is magnetized, a uniform magnetic field $\vec{B}$ (e.g. the earth's field) exerts a torque $\vec{\tau }_{{\rm B}} =\vec{M}(t)\times \vec{B}=M_{{\rm s}} \vec{e}_{{\rm M}} (t)\times \vec{B}$ on this dipole (with $M_{{\rm s}} $ being the saturation magnetization, $t$ the time and $\vec{e}_{{\rm M}} (t)$ the unit magnetization vector), and the net mechanical torque becomes:
\begin{equation} \label{GrindEQ__1_}
\vec{{\rm {\mathcal T}}}_{{\rm c}.m.} =\vec{\tau }_{{\rm F}s} +\vec{\tau }_{{\rm B}} (t)=\vec{R}\times \vec{F}_{{\rm s}} +\vec{M}(t)\times \vec{B}
\end{equation}
We introduce the following equation of motion for the magnetization in the coordinate system of the contact point:
\begin{equation} \label{GrindEQ__2_}
d\vec{M}(t)/dt=-\frac{\gamma \alpha }{(1+\alpha ^{2} )M_{{\rm s}} } \vec{M}(t)\times (\vec{M}(t)\times \vec{B})-\frac{1}{(1+\alpha ^{2} )} \vec{M}(t)\times \vec{\Omega }_{{\rm c.m.}} ,
\end{equation}
with the gyromagnetic ratio $\gamma $, the rolling angular velocity $\vec{\Omega }_{{\rm c}.{\rm m}} $ and the magnetic damping constant $\alpha $. Here, the first term accounts for the rotation of $\vec{M}$ towards the field due to $\vec{\tau }_{{\rm B}} $ \cite{LLG,Gilbert}, while the second term corresponds to the mechanical precession of $\vec{M}$ around a rolling axis.

First, we solve this set of equations analytically and numerically (see \cite{Supplemental}, part Methods) for $\vec{v}$, $\vec{M}(t)$ and $\vec{B}$ lying in the same plane $(xz)=\Pi $. If the amplitude of the mechanical torque $|\vec{\tau }_{{\rm Fs}} |$ surpasses that of the maximal possible magnetic torque $|\vec{\tau }_{{\rm B}}^{{\rm max}} |$, a sphere starts to roll in a usual way; that is, $d\vec{M}(t)/dt\ne 0$ and $\vec{{\rm {\mathcal T}}}_{{\rm c.m.}} \ne 0$ as shown in Fig. 2a. In contrast to a usual rolling, however, the sphere rolls non-harmonically due to $\vec{\tau }_{{\rm B}} $. If $|\vec{\tau }_{{\rm Fs}} |{\rm \; }\le {\rm \; }|\vec{\tau }_{{\rm B}}^{{\rm max}} |$, both torques may become compensated ($\vec{{\rm {\mathcal T}}}_{{\rm c.m.}} =0$) at $\sin [\angle (\vec{M}_{{\rm eq}} ,\vec{B})]=\sin [\theta _{{\rm eq}} ]=(RF_{{\rm s}} )/(M_{{\rm s}} B)$ (see \cite{Supplemental}, part B, Fig. 2b and the right panel of Fig. 1c). The magnetization relaxes towards $\vec{M}_{{\rm eq}} $ and remains at rest in rotational equilibrium. In the next step we allow for deviations of $\vec{v}$ and $\vec{B}$ from the $\Pi $ plane (see Fig. 2c-e). For $|\vec{\tau }_{{\rm Fs}} |{\rm \; }\le {\rm \; }|\vec{\tau }_{{\rm B}}^{{\rm max}} |$ the equilibrium magnetization will still relax to a direction minimizing the net torque $\vec{{\rm {\mathcal T}}}_{{\rm c.m.}} $, but this orientation will not belong to $\Pi $ anymore as shown in Fig. 2c,d. This, however, means that in addition to $\vec{\tau }_{{\rm Fs}} $ a torque $\vec{\tau }_{{\rm R}} =\vec{r}\times \vec{F}_{{\rm s}} $ emerges \cite{Cross,AA:EJP,Wang:AJP}, with $\vec{r}$ being a distance vector pointing from the axis $\vec{M}_{{\rm eq}} $ to the contact point (see Fig. 2e). If $\vec{{\rm {\mathcal T}}}_{{\rm c.m.}} =\vec{\tau }_{{\rm Fs}} +\vec{\tau }_{{\rm B}} $ vanishes, that is $d\vec{M}(t)/dt\to 0$ (Eq. (2)), the axis $\vec{M}_{{\rm eq}} $ becomes conserved and the sphere rotates up about it with angular momentum $\vec{L}_{{\rm R}} =I_{{\rm c.m.}} \vec{\Omega }_{{\rm R}} $ due to $\vec{\tau }_{{\rm R}} $ ($I_{{\rm c.m.}} $ is the inertia tensor). This emergent rotation is the essential contribution to the SR.
\begin{figure}
\includegraphics[width=0.75\linewidth]{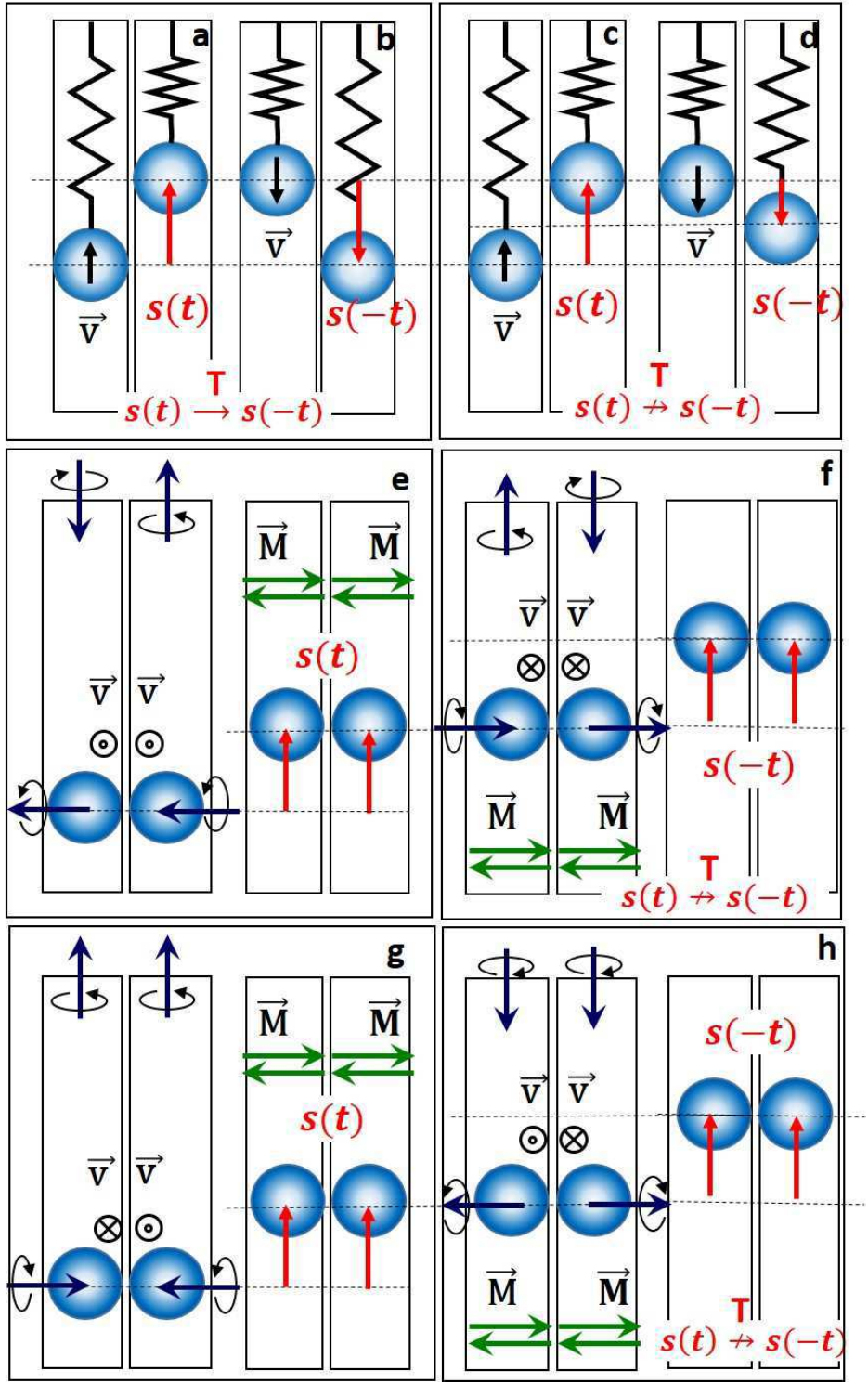}
	\caption{\textbf{Time-reversal symmetry breaking.} \textbf{(a-b)} Side-view of the forward-in-time ($s(t)$) and backward-in-time ($s(-t)$) trajectories of an ideal harmonic oscillator. Time-reversal symmetry is preserved. \textbf{(c-d)} Real harmonic oscillator with dissipation. Time-reversal symmetry is broken because $|s(t)|\ne |s(-t)|$, reversed order of events is preserved. \textbf{(e-h)} Side-view of the embodiment revealing a lifting force for different combinations of tubes' rotations corresponding to the forward-in-time \textbf{(e,g)} and the backward-in-time motion \textbf{(f,h)}. Black arrows indicate the angular momenta of the tubes and those of the spheres. Red arrows indicate trajectories. Green arrows show the allowed orientation of magnetization.}\label{fig:F4}
\end{figure}
Any rotation changes the trajectory of an object. Particularly, a spinning object acts as a gyroscope moving perpendicularly to the applied force $\vec{F}_{{\rm appl}} $ and obeying the dynamics of the gyroscope axis $\vec{\rho }$: $\frac{d\vec{\rho }}{dt} =\frac{R^{2} }{I_{{\rm c.m.}} \omega } (\vec{L}_{\rho } \times \vec{F}_{{\rm appl}} )$ \cite{Nash:PNAS2015}. There is, however, an important distinction between a standard gyroscope and a revolving magnet: the revolution axis is magnetic. The equation of motion becomes (with $\gamma $ being the gyromagnetic ratio):
\begin{equation} \label{GrindEQ__4_}
d\vec{M}_{{\rm eq}} /dt=\frac{\gamma {\rm \; }R^{2} }{I_{{\rm c.m.}} \Omega _{{\rm R}} } (\vec{L}_{{\rm R}} \times \vec{F}_{{\rm appl}} )
\end{equation}
This outcome is surprising. First, a magnetic sphere on an incline should revolve up spontaneously, without any external torque around $\vec{M}_{{\rm eq}} $ that is different from $\vec{\tau }_{{\rm B}} $, $\vec{\tau }_{{\rm Fs}} $ as well as their sum. Second, the time-reversal symmetry becomes broken as an action of the time operator \textit{T} on the left side of Eq. \eqref{GrindEQ__4_} $\frac{T(\vec{M}_{{\rm eq}} )}{T(t)} =\frac{T(-)}{T(-)} =T(+)$ differs from that on the right side  $T(\frac{R^{2} }{I_{{\rm c.m.}} \Omega _{{\rm R}} } )T(\vec{L}_{{\rm R}} )T(\vec{F}_{{\rm s}} )=T(+)T(-)T(+)=T(-)$ in contrast to a standard gyroscope with $\frac{T(\vec{\rho })}{T(t)} =\frac{T(+)}{T(-)} =T(-)$ on the left. Furthermore, the right side of Eq. \eqref{GrindEQ__4_} does not contain $\vec{M}_{{\rm eq}}$.

Thus, we have a unique situation: $\vec{M}_{{\rm eq}}$ defines the spatial alignment of $\vec{L}_{{\rm R}} $ but $\vec{M}_{{\rm eq}} $ and $\vec{L}_{{\rm R}} $ can be either parallel or antiparallel to one another, because the torque $\vec{\tau }_{{\rm R}} =\vec{r}\times \vec{F}_{{\rm s}} $ defining $\vec{L}_{{\rm R}} $ is independent of $\vec{M}_{{\rm eq}} $. Hence, the field reversal should result in the reversal of $\vec{M}_{{\rm eq}} $ but not in that of $\vec{L}_{{\rm R}}$ as shown in Fig. 2f. In other words, the reversal of $\vec{F}_{{\rm appl}}$ will result in the reversal of the SR trajectory, while the reversal of $\vec{B}$ will not. Hence, the SR leads to a kind of Lorentz force: for a given $\vec{F}_{{\rm appl}} $ and $\vec{B}$ a revolving sphere drifts to the right or to the left with $\vec{v}_{{\rm R}} (t)$. However, the sign of the Lorentz drift can be switched by the reversal of both, $\vec{B}$ or $\vec{F}_{{\rm appl}} $, while that of the SR-drift by the reversal of $\vec{F}_{{\rm appl}}$ only. This is the consequence of the time-reversal violation.

To check the conclusions as outlined above, we performed an experiment by letting hard magnetized NdFeB spheres to roll down an incline (see description and Movie S1-S2 \cite{Supplemental}). Initially, $\vec{M}$ (and the sphere) slowly precessed around $\vec{{\rm {\mathcal T}}}_{{\rm c.m.}} $ until $\vec{{\rm {\mathcal T}}}_{{\rm c.m.}} \to 0$ corresponding to $\vec{M}_{{\rm eq}} $ was reached. Then it rotated up around $\vec{M}_{{\rm eq}} $ and moved down an incline conserving $\vec{M}_{{\rm eq}} $ in agreement with the theoretical predictions. $\vec{M}_{{\rm eq}} $ was always collinear to $\vec{L}_{{\rm R}} $, but it was $\vec{M}_{{\rm eq}} \uparrow \downarrow \vec{L}_{{\rm R}} $ or $\vec{M}_{{\rm eq}} \uparrow \uparrow \vec{L}_{{\rm R}} $ depending on $\vec{F}_{{\rm appl}} $.  Furthermore, one can switch from clock-wise to counter-clock-wise angular velocity $\vec{\Omega }_{R} $ by changing the orientation of $\vec{B}$ with respect to the plane spanned by $\vec{F}_{{\rm appl}} $ and the surface normal. The gyroscopic drift of the revolving magnet can be seen in Movie S3 \cite{Supplemental}, where the reversal of $\vec{F}_{{\rm appl}} $ leads to the reversal of $\vec{v}_{{\rm R}} (t)$. In a reciprocal version of this experiment one can fix the revolving sphere by additional magnet $\vec{M}_{{\rm m}} $ and move the rolling surface instead of the sphere to achieve the SR (see Fig. 3a and Movie S4 \cite{Supplemental}).

In the next step we quantify the angular velocity $\vec{\Omega }_{{\rm R}} $ for a spontaneous rolling down an incline and a driven rolling as shown in Fig. 3a. For the rolling down an incline with $\vec{\Omega }_{{\rm R}} \bot \vec{v}$, an acceleration $a$ can be found analytically because of the simplification $\vec{v}_{{\rm R}} {\rm \parallel }\vec{v}$ (see \cite{Supplemental}, part C):
\begin{equation} \label{GrindEQ__5_}
a=\frac{5}{2} \frac{F_{{\rm s}} }{m} =\frac{5}{7} \; g\sin \beta
\end{equation}
Interestingly, it depends neither on $\theta _{{\rm eq}} $ nor on the mass $m$. Generally, $\vec{v}_{{\rm R}} \ne \vec{v}$ (Fig. 2f) and can be found numerically by deriving $\vec{M}_{{\rm eq}} $ from Eq. \eqref{GrindEQ__2_}, inserting the result into Eq. \eqref{GrindEQ__1_}, and solving Eqs. \eqref{GrindEQ__1_}-\eqref{GrindEQ__4_}. Fig. 3d shows $\vec{\Omega }_{{\rm R}} $ of a NdFeB sphere in three cases: \textbf{ I} corresponding to the set-up of Fig. 3a with linear velocity $\vec{v}=5$ m/s; \textbf{ II} corresponding to the rolling down an incline with $\beta =\pi /10$ and $\vec{B}=(-0.95,-0.22,0.22)10^{-5} $ T; and \textbf{ III} corresponding to the rolling down an incline with $\beta =\pi /10$ and $\vec{B}=(0,1,0)10^{-5} $ T. As one can see from this data $\vec{\Omega }_{{\rm R}} $ can be varied in a broad range by changing the applied force or inclination.

Now we switch to the SR in time-dependent fields. Let us consider two magnetic spheres, each put into a vertical non-magnetic tube. The tubes are placed close to one another and the spheres arrange themselves on internal sides of the tubes due to the magnetic attraction $\vec{F}_{{\rm m}}^{12} $ and $\vec{F}_{{\rm m}}^{21} $ \cite{Edwards:EJP} as shown in Fig. 3c,d and Fig. 4. If the tubes are rotated about their vertical axis due to $\vec{F}_{{\rm appl}} $, the spheres rotate initially together with the tubes. At a critical angle $\beta $, the sum of gravitational and magnetic forces overcomes the frictional force $\vec{F}_{{\rm s}} $ and the question is what happens then? A most intuitive answer is that the spheres return to their initial or to somewhat lower positions in response to $\vec{F}_{{\rm m}} +m\vec{g}$. On the other hand, at a critical angle $\beta $ the net torque $\vec{{\rm {\mathcal T}}}_{{\rm c.m}.} $ vanishes and, hence, we can expect the SR during the restoring motion. How will then the spin gyroscopes move? According to Eq. \eqref{GrindEQ__4_} the spheres should move upwards, which is counterintuitive. Our experiments, however, support the expectation of emerging revolution as well as that of a strong lifting force, defying gravity which tries to push the spheres downwards, and the magnetic interaction attracting the spheres in horizontal direction (see Fig. 3c and Movie S5 \cite{Supplemental}).

In time-reversal invariant systems, the equations of motion are invariant under the transformation $(q,\vec{p},t)\mapsto ^{T} (q,-\vec{p},-t)$ with $q$ being the coordinates, $\vec{p}$ the momentum and $t$ the time \cite{Baake:2006,Lamb:PhysD98}. In other words, the trajectory in reversed time should be a backward sequence of positions constituting the trajectory in forward time. To check this, one reverses the momentum $\vec{p}$ and looks for the corresponding trajectory. If one reverses the rotational momentum of the tubes, the spheres will not go downwards. They will repeatedly move upwards to any tube height (Fig. 4, Movie S5\cite{Supplemental}) breaking the time-reversal symmetry. Importantly, this symmetry breaking is neither local, like that of a Lorentz force, nor dissipation-driven. Indeed, the trajectory of a charge due to the Lorentz force becomes time-reversal invariant if the direction of magnetic field is reversed, because $\vec{B}\mapsto ^{T} -\vec{B}$ (see \cite{Supplemental}, part D and Fig. S1 \cite{Supplemental}). The only way to force the spheres moving downwards is to reverse the gravitational force. This operation is, however, forbidden as the forces are even under time-reversal ($\vec{F}\mapsto ^{T} \vec{F}$). Dissipation is also a known source for the violation of time-reversal symmetry as shown in Fig. 4a-d. In this case, however, the trajectory's length changes while the reversability of time events is not affected. In case of SR the reverse tape effect is impossible as shown in Fig. 4e-h: the time reversal results in a new trajectory. While friction is one of reasons for both phenomena: the SR and the energy dissipation, the latter is neither the reason for the SR, nor for the described time-reversal symmetry breaking. Rather, this symmetry violation stems from the emergent revolving up of the magnet and subsequent curved trajectory as explained in \cite{Supplemental}, part E-F. In case of tubes, the SR is achieved due to combination of magnetic attraction, friction and gravitation. It is, however, important that $m\vec{g}$ does not belong to the $\Pi $ plane defined by $\vec{F}_{{\rm appl}} $and $\vec{N}$. If $m\vec{g}\in \Pi $, e.g. the tubes lie on a horizontal surface, the SR does not appear (see \cite{Supplemental}, part G and Fig. S3 \cite{Supplemental}). However, already tiniest deviation from the horizontality ensures the SR.  The upper limit of the lifting force can be approximated by $F_{{\rm lift}} (r_{12} )\approx F_{{\rm m}} (r_{12} )-kF_{{\rm m}} (r_{12} )\cos \beta -mg$ with $k$ being the friction coefficient. As the rolling friction $F_{{\rm r}S} $ is tiny (0.05-0.07 for metal/plastic interfaces), $F_{{\rm lift}} $ can reach significant values.

To conclude, we presented a novel magnetomechanical effect consisting of rotating up a magnet and subsequent gyroscopic motion, thereby breaking time-reversal symmetry. This phenomenon offers a variety of promising applications in different fields of science and engineering including the delivery of magnetic (nano)particles. Particularly, the SR effect can be used to achieve controllable translation of objects or magnetic particles in any direction on vertical or horizontal surfaces as shown in Movie S6 \cite{Supplemental}. The advantage of this motion is the absence of direct contact between the tubes and the absence of any kind of guides increasing the weight and complexity of the system. Furthermore, the SR effect can be used for effective interconversion between rotational and translational motion that is important for linear or angle motors as shown in Movie S4 \cite{Supplemental}. The advantage of this kind of conversion is the absence of any kind of gears and versatile possibilities of switching the rotational sense. Additionally, the lifted magnets can be used for energy storage and its later harvesting using magnetic induction. An array of revolving magnets can also be utilized as information storage element. Thus, the SR effect will change our perspectives of existing magnetic phenomena and open up new technological possibilities for energy storage, energy interconversion and robotics.

{\em Acknowledgements.}
We gratefully acknowledge funding from the Cluster of Excellence  'Advanced Imaging of Matter' (EXC 2056 - project ID 390715994) of the Deutsche Forschungsgemeinschaft (DFG).

%{\em Author Contributions.}

%\bibliography{ICE}

\begin{thebibliography}{100}

\bibitem{Du:2021}L. Du, T. Hasan, A. Castellanos-Gomez, G.-B. Liu, Y. Yao, C. N. Lau, Z. Sun, Engineering symmetry breaking in 2D layered materials, Nat. Rev. Phys. \textbf{3}, 193 (2021)
\bibitem{Smej:CA2020} L. Šmejkal, R. Gonz\'ales-Hern\'andez, T. Jungwirth, J. Sinova, Crystal time-reversal symmetry breaking and spontaneous Hall effect in collinear antiferromagnets, Sci. Advances \textbf{6}, 1-9 (2020).
\bibitem{LL} L. D. Landau and E. M. Lifshitz, Mechanics (Pergamon, New York, 244 p., 1969).
\bibitem{Matsuo:PRL11} M. Matsuo, J. Ieda, E. Saitoh, and S. Maekawa, Effects of Mechanical Rotation on Spin Currents, Phys. Rev. Lett. \textbf{106}, 076601 (2011)
\bibitem{Matsuo:JPSJ17} M. Matsuo, E. Saitoh, S. Maekawa, Spin-Mechatronics, J. Phys. Soc. Jpn. \textbf{86}, 011011 (2017)
\bibitem{Heims} S. P. Heims and E. T. Jaynes, Theory of Gyromagnetic Effects and Some Related Magnetic Phenomena. Rev. Mod. Phys. 34, 143-164 (1962).
\bibitem{Barnett} S. J. Barnett, Magnetization by rotation, Phys. Rev.  \textbf{6}, 239 (1915).
\bibitem{EinsteinHaas} A. Einstein and W. J. de Haas, Experimenteller Nachweis der Ampereschen Molekularstr\"ome  (Experimental Proof of Amp\'ere's Molecular Currents), Verh. Dtsch. Phys. Ges.  \textbf{17}, 152 (1915)
\bibitem{Supplemental} Supplemental Information.
\bibitem{ChemPhysChem} E. Y. Vedmedenko and N. Mikuszeit, Multipolar Ordering in Electro- and Magnetostatic Coupled Nanosystems, Chem. Phys. Chem. \textbf{9}, 1222 (2008)
\bibitem{Edwards:EJP} B. F. Edwards and J. M. Edwards, Dynamical interactions between two uniformly magnetized spheres, Eur. J. Phys. \textbf{38}, 015205 (2017)
\bibitem{LLG} L. D. Landau, E.M. Lifshitz, On the Theory of the Dispersion of Magnetic Permeability in Ferromagnetic Bodies, Physik. Zeits. Sowjetunion \textbf{8}, 153 (1935)
\bibitem{Gilbert} T. Gilbert, A Phenomenological Theory of Damping in Ferromagnetic Materials, Magnetics, IEEE Trans. \textbf{40}, 3443 (2004)
\bibitem{Cross} R. Cross, Precession of a Spinning Ball Rolling Down an Inclined Plane, The Physics Teacher \textbf{53}, 217 (2015)
\bibitem{AA:EJP} C. Aghamohammadi and A. Aghamohammadi, Slipping and Rolling on an inclined Plane, Eur. J. Phys. \textbf{32}, 1049 (2011)
\bibitem{Wang:AJP} X. Wang, Trajectory of a projectile on a frictional inclined plane, Am. J. Phys. \textbf{82}, 764 (2014)
\bibitem{Nash:PNAS2015} L. M. Nash, D. Kleckner, A. Read, V. Vitelli, A. M. Turner, and W. T. M. Irvine, Topological mechanics of gyroscopic metamaterials, PNAS \textbf{112}, 14495 (2015).
\bibitem{Lamb:PhysD98} J.S.W. Lamb and J.A.G. Roberts, Time-reversal symmetry in dynamical systems: A survey, Physica D \textbf{112},1 (1998)
\bibitem{Baake:2006} M. Baake and J.A.G. Roberts, The structure of reversing symmetry groups, Bulletin of the Australian Mathematical Society \textbf{73}, 445 (2006).






\end{thebibliography}

\end{document}